\documentclass[11pt,preprint]{aastex}

\usepackage{graphicx}
\usepackage{subfigure}
\usepackage{subfloat}
\usepackage{amssymb}
\usepackage{amsmath}
\usepackage{multirow}
\usepackage{dcolumn}
\usepackage{bm}

\shorttitle{Collapsed Cores in Globular Clusters}
\shortauthors{}

\begin{document}

\title{Dynamo model for the inverse transfer of magnetic energy in a nonhelical decaying magnetohydrodynamic turbulence}

\author{Kiwan Park}
\affil{Department of Physics, UNIST, Ulsan, 689798, Korea; pkiwan@unist.ac.kr}

\begin{abstract}
The inverse cascade of magnetic energy occurs when helicity or rotational instability exists in the magnetohydrodynamic (MHD) system. This well known phenomenon has been considered as a basis for the large scale magnetic field in universe. However nonhelical magnetic energy in a decaying MHD system also migrates toward the large scale, which holds vital clues to the origin of large scale magnetic field in a quiescent astrophysical system. Zeldovich's rope dynamo model is considered as a basic and symbolistic model of magnetic field amplification. However, the rope model assuming specific external forces like buoyancy or Coriolis force is not appropriate for a decaying turbulent system without any external force. So we suggest a new dynamo model based on magnetic induction equation excluding a forcing source. This model shows the expansion and growth of magnetic field (flux) is basically the redistribution of energy in the system. The transfer of magnetic energy is in fact a successive induction of magnetic field resulted from the interaction between the fluid motion and seed magnetic field. We also discuss about an analytic theorem based on the scaling invariant MHD equation.
\end{abstract}


\section*{Introduction}
Magnetic field is a ubiquitous phenomenon in universe which is full of conducting fluids (plasmas). The interaction between magnetic field and plasma is known to play a crucial role in the evolution of celestial objects like pulsars, jets, galaxy, or GRBs etc. So better comprehension of the rudimentary interaction causing the MHD phenomena will more clearly explain the past and present universe, and how the universe will evolve. However since the interaction between the magnetic field and conducting fluid is basically a nonlinear process, the intuitive understanding of MHD phenomena without its exact solution is not easy nor reliable. At the moment the general solution of MHD equation is not yet known. Only a few approximate (stochastic) solutions \citep{Kraichnan and Nagarajan 1967, Kazantsev 1968, Pouquet et al 1976, Krause and Radler 1980} are available in a limited way, which means our comprehension is also restricted as much. So computational simulation is widely used as an alternative method. However the exact interpretation of simulation, which is the numerical calculation of MHD equation, requires precise understanding of the analytic theories on MHD. So the investigation of a theory-based intuitive model with the numerical simulation is a practical and helpful way to get the point of MHD phenomena.\\

\noindent We briefly introduce couple of models on the origin of magnetic field in the early universe. We show simulation results for the amplification of (nonhelical) magnetic field in a decaying MHD system. Then we discuss about a theory derived from the scaling invariant MHD equation \citep{Olesen 1997}. Finally we will introduce an intuitive dynamo model for the general mechanism of magnetic energy transfer in plasma. They give us clues to the origin of magnetic field in the quiescent astrophysical system.\\

\noindent For the origin of magnetic field roughly two hypotheses are accepted: primordial (top-down) and astrophysical (bottom-up) model. Primordial hypothesis supposes magnetic field could be generated as the conformal invariance of electromagnetic field was broken during the inflationary period of early universe \citep{Turner and Widrow 1988}. After the expansion, magnetic field could be successively generated through the cosmic phase transition such as electroweak phase transition (EWPT) or quantum chromodynamic transition (QCD) from quark to hardron \citep{Grasso and Rubinstein 2001, Subramanian 2015}. The magnitude of generated magnetic field are thought to have been $\mathrm{B}_0\sim 10^{-62}$G on a 1 Mpc comoving scale (during inflation), $\sim 10^{-29}$G (EWPT), and $\sim 10^{-20}$G (QCD) on a 10 Mpc scale \citep{Sigl and Olinto 1997}. The correlation lengths of these seed fields were limited by the scale of particle horizon: $\sim 1$cm (EWPT) and $\sim 10^4$ cm (QCD). On the other hand astrophysical hypothesis, posterior to the primordial inflation, suggests that the seed field was generated by the plasma effect in the primeval astrophysical objects like galaxies or clusters. The strengths of magnetic seed fields in these models are inferred to be in the range between $B_0$$\sim 10^{-21}$G (Biermann battery effect) and $\sim 10^{-19}$G (Harrison effect).\\

\noindent However, whether the seed magnetic fields were originated from primordial or astrophysical model, the inferred strengths are too weak for the currently observed magnetic fields ($\sim\mu$G). Also the inferred correlation length, which should be limited to the particle horizon at that time, is too tiny compared with the that of present magnetic field in space. Definitely the initial seed fields must have been amplified \citep{Cho 2014}. This process, called dynamo, is essentially the redistribution of energy in the MHD system. If the magnetic energy is cascaded toward large scale with its intensity increasing, this is called large scale dynamo (LSD, \cite{Brandenburg 2001}, \cite{Blackman and Field 2002}). In contrast if the energy is cascaded toward small scale, the process is called small scale dynamo (SSD, \cite{Kulsrud and Anderson 1992}). Also if the growth rate of magnetic field depends on the magnetic resistivity, the process is called slow dynamo; otherwise, it is called fast dynamo. The migration and increase of magnetic field are influenced by many factors whose critical conditions are still in debate. But, we will not discuss about the criteria at this time. We focus on the qualitative mechanism of energy transfer in the MHD system.\\

\noindent As mentioned dynamo is essentially energy redistribution. In the turbulent system the direction of energy transfer is related to the conservation or minimization of system variables. For example in (quasi) two dimensional ideal hydrodynamics the kinetic energy $u^2$ and enstrophy $\omega^2=(\nabla \times \mathbf{u})^2$ are conserved. Then using $\frac{d}{dt}u^2\sim\frac{d}{dt}\int^\infty_0 E_V(k,t)dk=0$ and $\frac{d}{dt}\omega^2\sim \frac{d}{dt}\int^\infty_0 k^2 E_V(k,t) dk=0$ we can trace the migration of kinetic energy like below \citep{Davidson 2004}:
\begin{eqnarray}
\frac{d}{dt}\langle k\rangle &=&\frac{d}{dt}\bigg(\frac{\int^\infty_0kE_V(k,t)dk}{\int^{\infty}_0E_V(k,t)dk}\bigg)
=\frac{\frac{d}{dt}\int^\infty_0kE_V(k,t)dk}{\int^{\infty}_0E_V(k,t)dk}.
\label{eqs1}
\end{eqnarray}
With the definition of initial centroid of $\langle E_V\rangle$ by $k_c=\int kE(k)dt/\int E(k)dt$, this result can be rewritten like
\begin{eqnarray}
\frac{d}{dt}\langle k\rangle =-\frac{1}{k_{c}}\frac{\frac{d}{dt}\int^\infty_0 (k-k_{c})^2 E_V(k,t)dk}{\int^\infty_0E_V(k,t)dk}.
\label{eqs2}
\end{eqnarray}
Since the naturally spreading energy spectrum $E_V(k,t)$ makes $\int^\infty_0 (k-k_{c})^2 E_V(k,t) dk$ increase, the peak of $E_V$ retreats toward the large scale with time; i.e., the inverse cascade of $E_V$. In three dimensional case where the enstrophy is not conserved, $E_V$ migrates toward the smaller scale which has larger damping effect. But with the kinetic helicity $H_V$ ($=\langle \mathbf{u}\cdot \omega \rangle,\, \omega=\nabla \times \mathbf{u}\sim \mathbf{u}$), which is a conserved variable in the ideal hydrodynamic system, the energy can be inversely cascaded with the triad interaction among eddies \citep{Biferale et al 2012}.

In the MHD system without helicity, the kinetic and magnetic energy cascade toward small scale. But if the MHD system has a nontrivial kinetic helicity $H_V$, or magnetic helicity $H_M\,(=\langle \mathbf{A}\cdot \mathbf{B} \rangle$, $\mathbf{B}=\nabla \times \mathbf{A}\sim \mathbf{A}$), magnetic energy is cascaded inversely \citep{Brandenburg and Subramanian 2005}. This phenomenon can be derived using the conservation of energy and helicity, which is also related to the minimization of energy in the system. Since the MHD system converges into the absolute equilibrium state, the system variables can be generally described by Gibbs functional composed of ideal invariants: $\langle u^2+B^2\rangle$, $\langle \mathbf{A}\cdot \mathbf{B}\rangle$, and $\langle \mathbf{u}\cdot \mathbf{B}\rangle$. The ensemble average of a N-dimensional system variable is represented by the partition function `$Z$' and the generalized coordinate `$p$', `$q$': $\langle A \rangle = \int\, \rho\, A(p,\,q)\,dp^{3N}\,dq^{3N}$, $\rho= Z^{-1}exp\,[-W]$, $Z= \int\, exp\,[-W]\,dp^{3N}\,dq^{3N}$, $W=\lambda_1 E+\lambda_2 \textbf{A}\cdot \textbf{B}+\lambda_3 \textbf{u}\cdot \textbf{B}$. However, the system invariant itself does not always guarantee the inverse cascade. What determine the direction of cascade are the structures of ideal invariants and Lagrangian multipliers: $\lambda_1$, $\lambda_2$, and $\lambda_3$ \citep{Frisch et al. 1975}. From a fluid point of view, in addition to the statistical concept, we can explain the inverse cascade of $E_M$ using the mean field theory \citep{Blackman and Field 2002, Park and Blackman 2012a, Park and Blackman 2012b}. The profiles of large scale magnetic energy $\overline{E}_M$ and helicity $\langle \overline{\mathbf{A}}\cdot \overline{\mathbf{B}} \rangle$ in Fourier space ($k=1$) are represented like below \citep{Park 2014}:
\begin{eqnarray}
\frac{\partial }{\partial t}\overline{E}_M
&=&\alpha\langle \overline{\mathbf{A}}\cdot \overline{\mathbf{B}} \rangle-2(\beta+\eta)\overline{E}_M,\label{Em with alpha coefficient}\\
\frac{\partial }{\partial t}\langle \overline{\mathbf{A}}\cdot \overline{\mathbf{B}} \rangle
&=&4\alpha \overline{E}_M-2(\beta+\eta)\langle \overline{\mathbf{A}}\cdot \overline{\mathbf{B}} \rangle \label{Hm with alpha coefficient},
\end{eqnarray}
where
\begin{eqnarray}
\alpha=\frac{1}{3}\int^t\big(\langle {\bf j}\cdot {\bf b}\rangle-\langle {\bf u}\cdot {\bf\omega}\rangle\big ) d \tau,\,\, \beta=\frac{1}{3}\int^t\langle u^2\rangle\,d\tau.
\label{Alpha coefficient}
\end{eqnarray}
Here, $\langle {\bf j}\cdot {\bf b}\rangle$, $\langle {\bf u}\cdot {\bf\omega}\rangle$, and $\langle u^2\rangle$ are respectively small scale current helicity, kinetic helicity, and kinetic energy. These equations indicate the energy in small scale is transferred to the large scale through the small scale helical field ($\alpha\sim\langle {\bf j}\cdot {\bf b}\rangle-\langle {\bf u}\cdot {\bf\omega}\rangle$), and the dissipation of large scale magnetic field is related to magnetic resistivity and small scale kinetic energy ($\beta\sim \langle u^2\rangle$). This equation is valid for both a driven and decaying helical MHD system.\\

\noindent Also the cross field diffusive flow \citep{Ryu and Yu 1998} or Bohm diffusion effect \citep{Lee and Ryu 2007} was forced to increase the flux of magnetic field. Their concepts are somewhat different from the typical dynamo theory, but include the fundamental question on the dynamo.\\

\noindent In a driven MHD system the saturated field profile is decided by the injection scale, external driving force, and the intrinsic properties like viscosity \& magnetic resistivity, not by the initial conditions whose effect disappears within a few simulation time steps \citep{Park 2013, Park 2014}. In contrast in a free decaying MHD system the profile of decaying field is determined by the initial conditions and selective decaying speed besides other intrinsic properties. Especially the selective decaying speed due to the different spatial derivative order in the `system invariants' arouses academic interest besides many applications: the life time of star-forming clouds, measurement of galactic magnetic field, or fast magnetic reconnection \citep{Biskamp 2008}. However, we should note the free decay does not mean the interaction between magnetic field and plasma is forbidden. While the total energy decays with time, the magnetic field still interacts with the fluid to migrate among eddies. Whether or not the MHD system is driven, the helical $E_M$ is inversely cascaded as long as `$\alpha$' in Eq.(\ref{Em with alpha coefficient}) is predominant over the dissipation effect. Moreover the magnetic helicity in large scale, the statistical correlation between different components of magnetic fields, is resilient to the turbulent diffusion so that the decay speed slows down \citep{Blackman and Subramanian 2013}. On the other hand the inverse transfer of $E_M$ in a decaying system without helicity or shear is contradictory to the typical MHD dynamo theory \citep{Olesen 1997, Shiromizu 1998, Zrake 2014, Brandenburg et al 2014}. This phenomenon implies the energy released from the past events like supernovae could be a source of large scale magnetic field structure in a quiescent astrophysical object which does not have a significant driving source.\\

\noindent Usually the amplification of magnetic field (flux) due to the fluid motion is explained by Zeldovich's rope model `stretch, twist, fold, and merge' \citep{Zeldovich et al 1983}. However, since the model assumes the influence of external force, it is not appropriate to the growth of large scale magnetic field in a decaying turbulence.\\

\section{Simulation and method}
We used $\mathrm{PENCIL}$ $\mathrm{CODE}$ \citep{Brandenburg 2001} for the weakly compressible fluid in a periodic box $(8 \pi^3)$. MHD equations are basically coupled partial differential equations composed of density `$\rho$', velocity `$\mathbf{u}$', and vector potential `$\mathbf{A}$' (or magnetic field $\mathbf{B}=\nabla \times \mathbf{A}$).
\begin{eqnarray}
\frac{D \rho}{Dt}&=&-\rho {\bf \nabla} \cdot {\bf u}\label{density conservation},\label{continuity equation for pencil code}\\
\frac{D {\bf u}}{Dt}&=&-{\bf \nabla} \mathrm{ln}\, \rho + \frac{1}{\rho}(\nabla\times{\bf B})\times {\bf B}+\nu\big({\bf \nabla}^2 {\bf u}+\frac{1}{3}{\bf \nabla} {\bf \nabla} \cdot {\bf u}\big)\label{momentum equation for pencil code},\\
\frac{\partial {\bf A}}{\partial t}&=&{\bf u}{\bf \times} {\bf B} -\eta\,{\bf \nabla}{\bf \times}{\bf B}+\mathbf{f}.\label{magnetic indution equation for pencil code}
\label{MHD equations in the pencil code}
\end{eqnarray}
Here $D/Dt(=\partial / \partial t + {\bf u} \cdot {\bf \nabla}$) is Lagrangian time derivative to be calculated following the trajectory of fluid motion. In simulation we used 0.015 and $2\times 10^{-5}$ for the kinematic viscosity `$\nu$' and magnetic diffusivity `$\eta$' respectively. This setting is to realize the large magnetic prandtl number $Pr_M$ (=$\nu/\eta$) in the early universe as similar as we can. The function $\mathbf{f}(\mathbf{x},t)$ generates the random nonhelical force with the dimension of electromotive force (EMF, $\mathbf{u}\times \mathbf{B}$). The nonhelical magnetic energy maximizes Lorentz force $\mathbf{J}\times \mathbf{B}$ so that the fluid motion in plasma is efficiently excited. Turning on and off the function ($0<t<1$, simulation time unit), we imitate a celestial system which was left decaying after it had been driven magnetically by an event in the past.

\section{Results}
\begin{figure}
\centering{
  {
   \subfigure[]{
     \includegraphics[width=8cm]{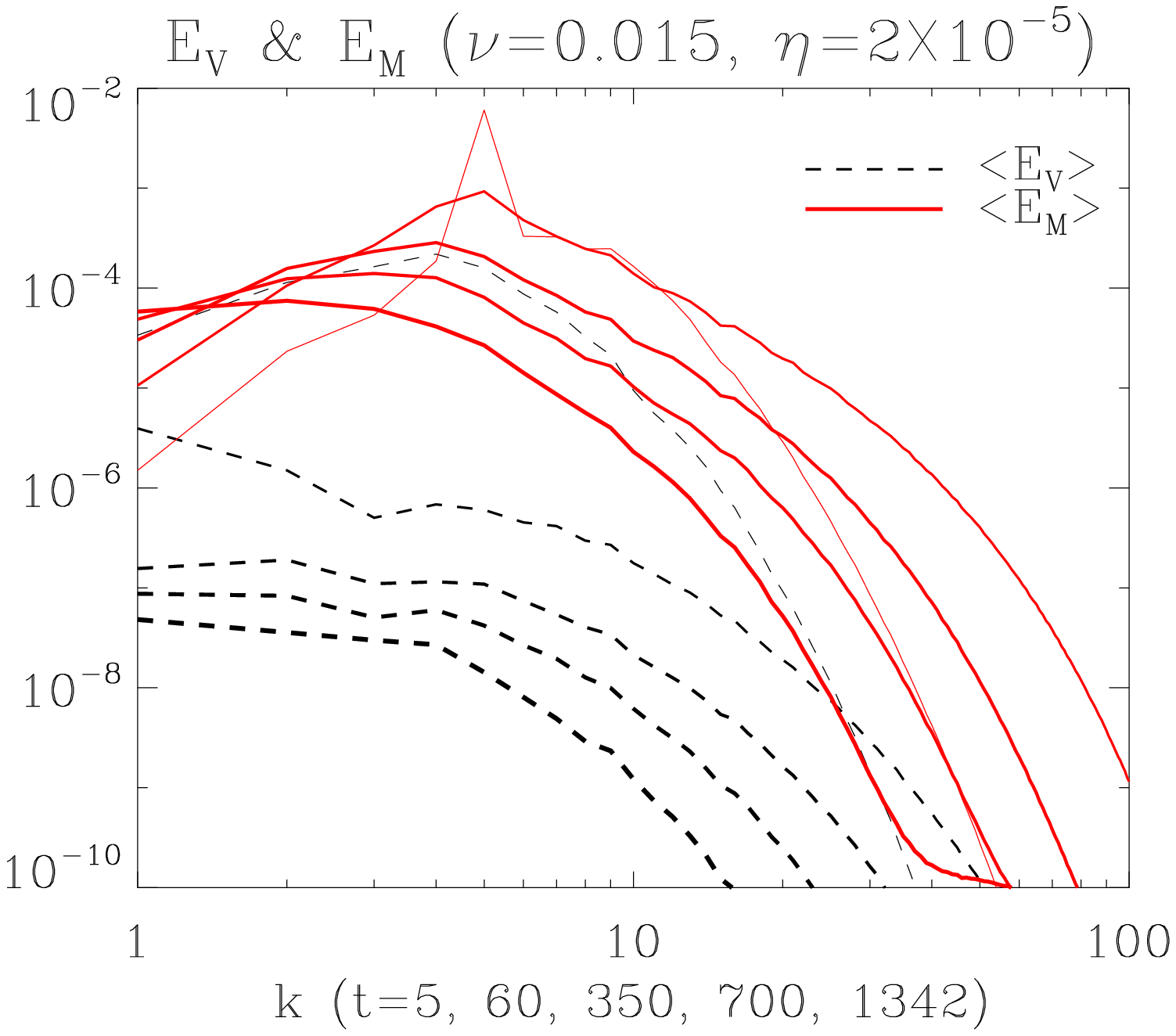}
     \label{f1}
   }\hspace{-10mm}
   \subfigure[]{
     \includegraphics[width=8cm]{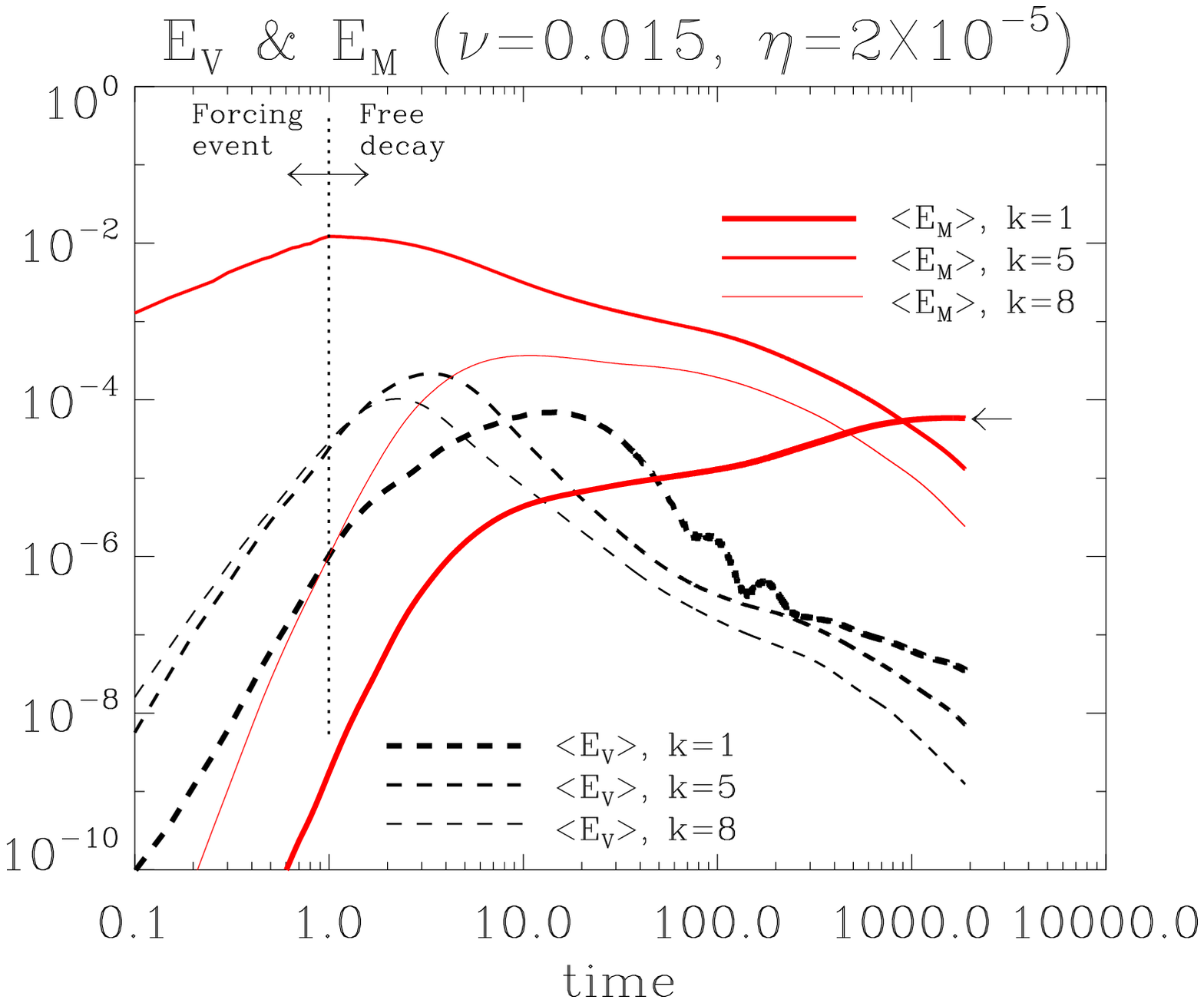}
     \label{f2}
     }
  }
\caption{(a) Free decaying $E_V$ and $E_M$ spectrum after nonhelical forcing. Vertical axis means the energy, and horizontal axis means the wave number. (b) The time evolution of $E_V$ and $E_M$. Vertical axis is the energy, and horizontal one is time. The lines of $k$=1, 5, and 8 represent the large, injection, and small scale energy spectrum respectively. The plot shows the increasing energy in large and small scale is in fact the transferred one from the injection scale at $k=5$.}
}
\end{figure}

Fig.\ref{f1} shows the evolution of kinetic and magnetic energy spectrum in a decaying MHD system after the initial nonhelical magnetic forcing ($0<t<1$). Black dotted line indicates kinetic energy spectrum $E_V$, and red solid one indicates magnetic energy spectrum $E_M$. The increasing thickness of line means the lapse of time from $t=5$ (thinnest) to $t=1342$ (thickest). The overall $E_V$ and $E_M$ decrease after the forcing stops, but each spectrum does not monotonically decay. For $5<t<60$ $E_V$ in the small scale ($k>\sim 25$) grows before it turns into the decay mode at $t\sim60$, and the lagging decrease of $E_V$ in the large scale gives some hints of the influence of eddy turnover time `$\tau$'. The profile of $E_M$ shows the similar, but somewhat different pattern. $E_M$ in small scale grows faster than that of large scale, surpasses its initial magnitude, and then begins to decrease at $t\sim60$. In contrast, $E_M$ in large scale keeps growing far longer before it eventually decays. The overall evolution of magnetic field lags behind that of velocity field, which is a typical feature of a large $Pr_M$ MHD system\\

\noindent Fig.\ref{f2} in fact has the same information as that of Fig.\ref{f1}. But it shows the important features of a decaying nonhelical field more clearly. After the event stops at $t\sim1$, both $E_V$ and $E_M$ grow temporarily before they decay. The onsets of large scale $E_M$ and $E_V$ lag behind those of small scale energy but keep growing longer, which implies the influence of $\tau$, $Pr_M$, and energy supply on the evolution of large scale field. The advection term `$-\mathbf{u}\cdot \nabla \mathbf{u}$' or pressure  `$-\nabla p$' transfers $E_V$ to make the system homogeneous and isotropic without the help of $E_M$. But for the transfer of $E_M$, $E_V$ is a prerequisite as the source term `$\nabla \times \langle \mathbf{u}\times \mathbf{b}\rangle$' in the magnetic induction equation shows. The energy in the kinetic eddy is non locally transferred to the magnetic eddy through $\mathbf{B}\cdot \nabla \mathbf{u}$, and the magnetic energy is locally transferred through -$\mathbf{u}\cdot \nabla \mathbf{B}$. Therefore the long lasting $E_V$ in large scale can induce the migration of $E_M$ toward the large scale, which also keeps longer due to the proportionally increasing eddy turnover time ($\sim 1/k$) and decreasing magnetic diffusivity ($\sim k^2$). The inverse transfer of $E_M$ without helicity or shear deviates from the principle of typical dynamo theory. However the seemingly inconsistent phenomenon is true and implies the fundamental principle of energy migration in the magnetized plasma system. \cite{Olesen 1997} suggested the (inverse) transfer of energy be an essential phenomenon in the scaling invariant (self-similar) MHD equations, and the direction of energy transfer should be decided by the initial distribution of given energy `$E(k,0)\sim k^q$'. The self similarity theorem shows that energy is inversely transferred if $q<-3$; otherwise, the energy cascades forward. The inverse transfer of $E_M$ in a decaying MHD system implies that an event, which emitted electromagnetic energy in the past, can be an origin of large and small scale magnetic field observed in the present universe. As an application of this plot, an observer at `$t\sim 1000$' can find large and small scale $E_V$ \& $E_M$ without any driving source nor helicity.\\

\begin{figure}
\centering{
  {
   \subfigure[]{
     \includegraphics[width=8cm]{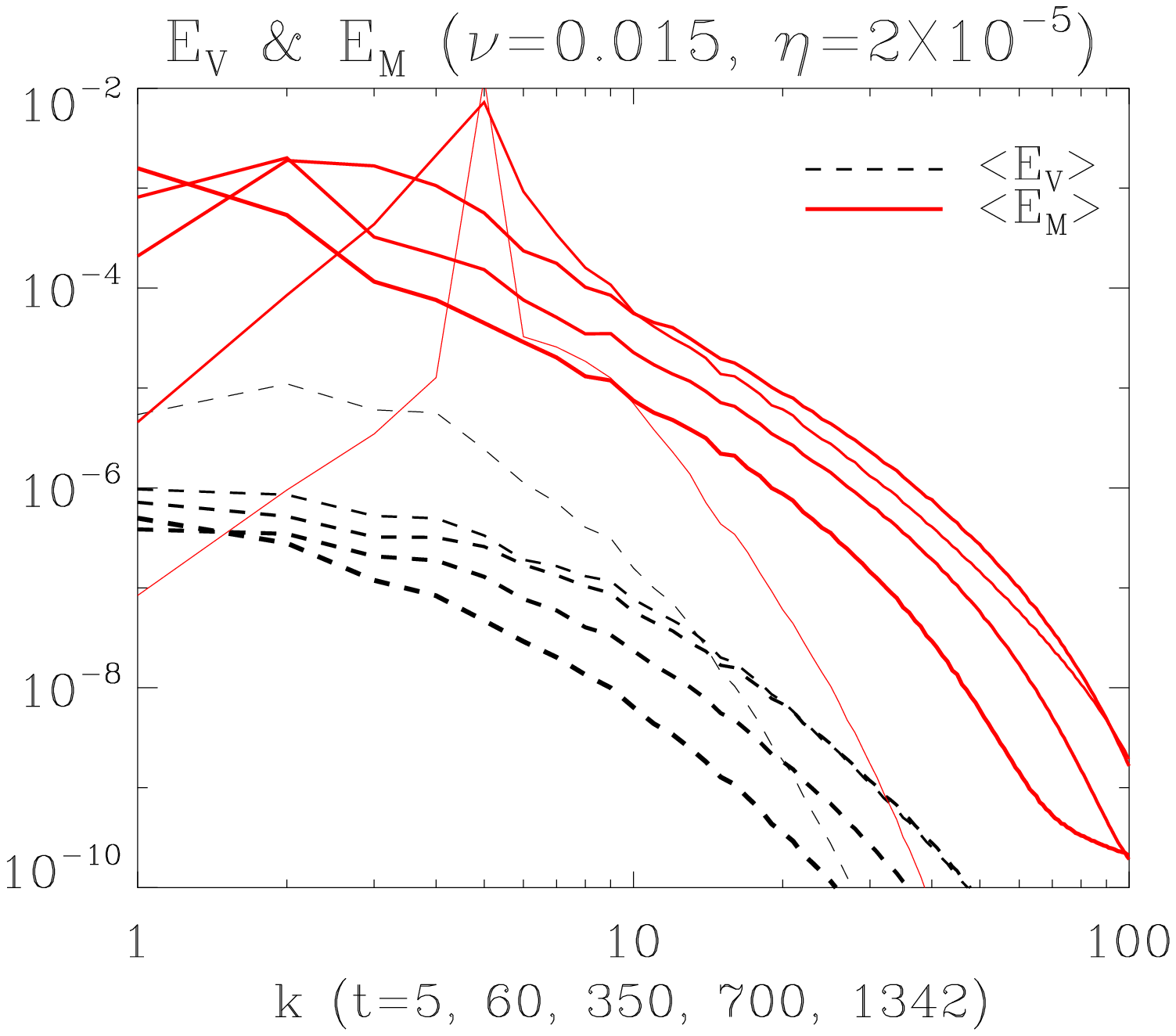}
     \label{f1a}
   }\hspace{-10mm}
   \subfigure[]{
     \includegraphics[width=8cm]{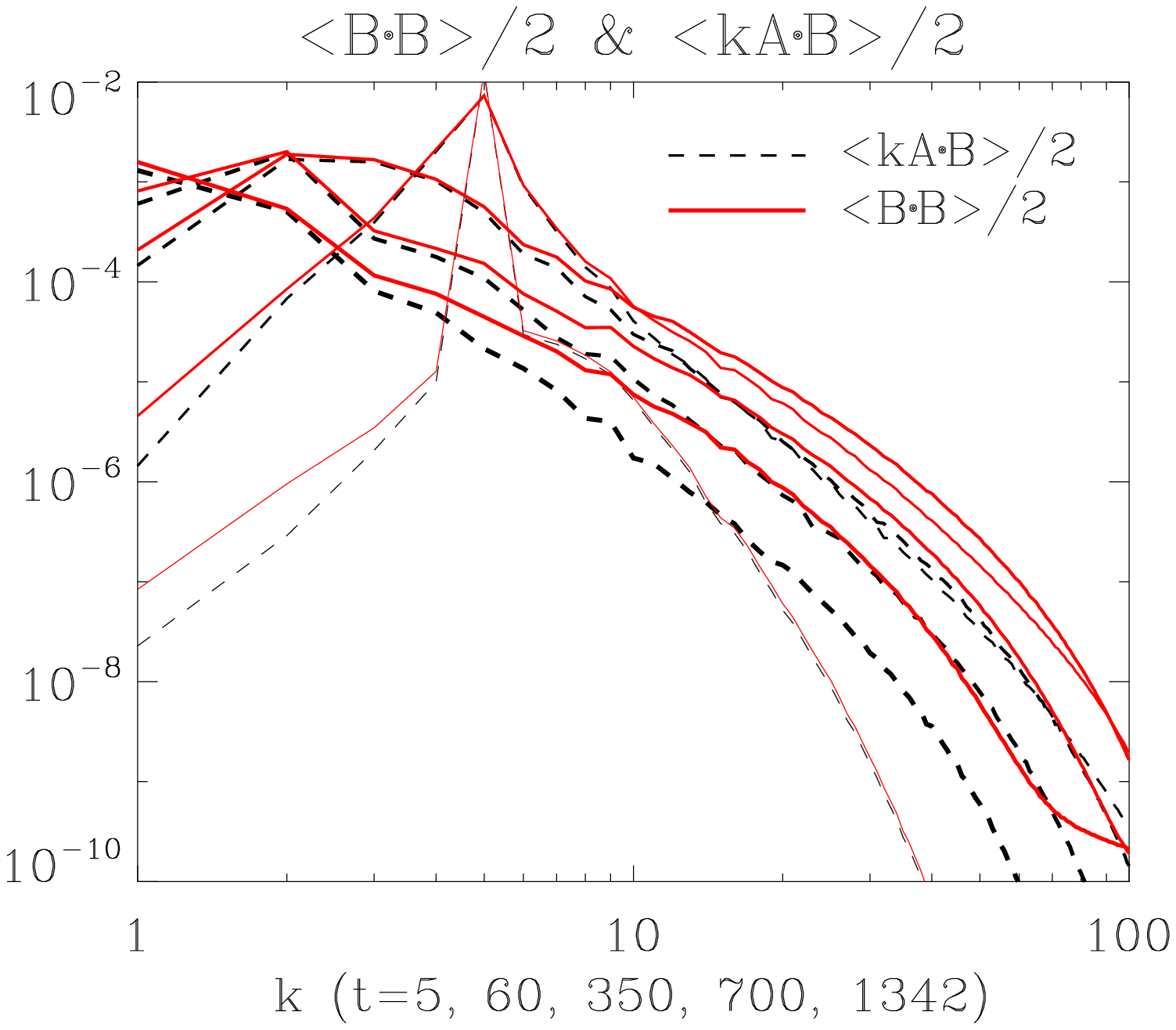}
     \label{f1ab}
   }\hspace{-10mm}
   \subfigure[]{
     \includegraphics[width=8cm]{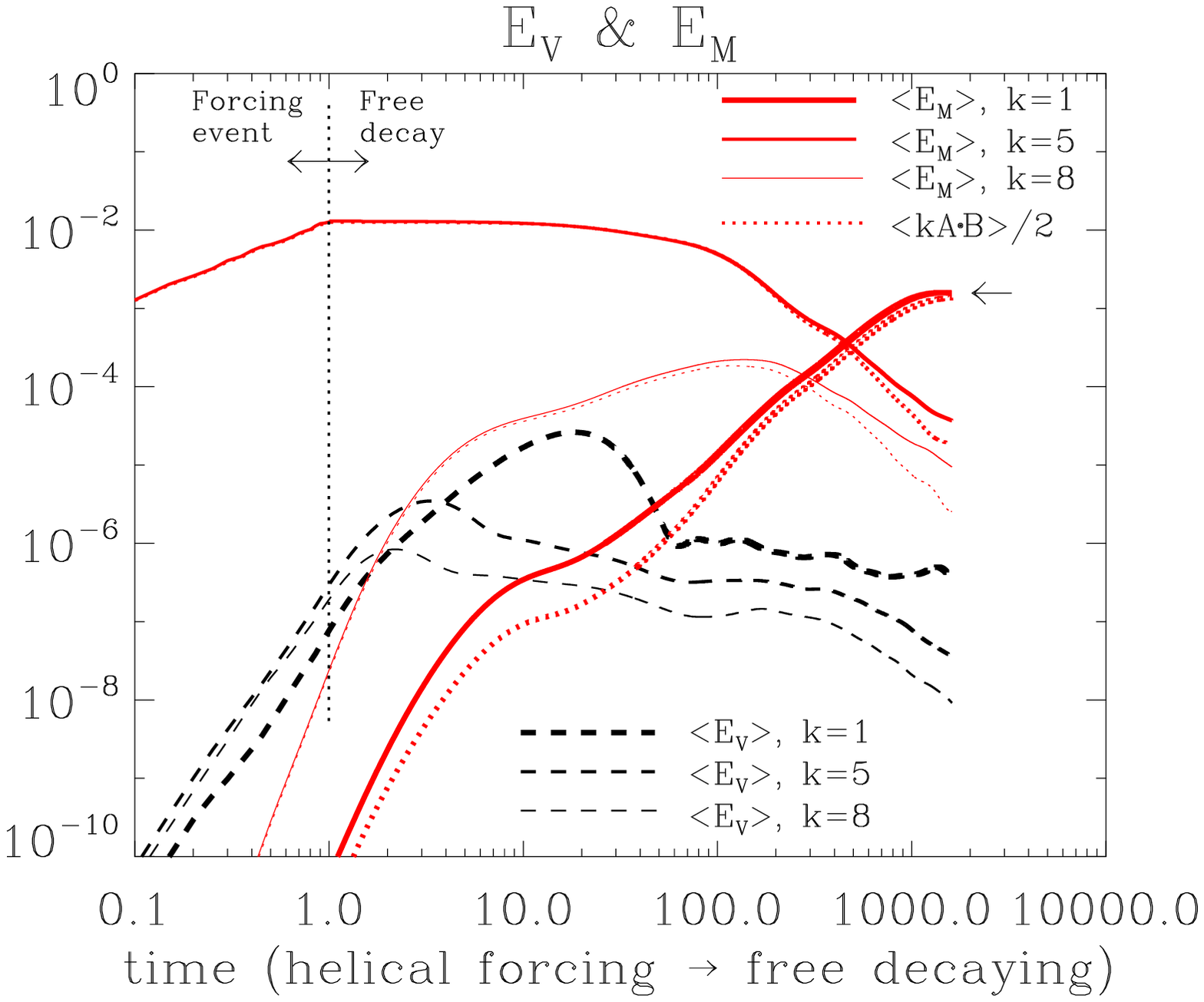}
     \label{f2a}
     }\hspace{-10mm}
   \subfigure[]{
     \includegraphics[width=8cm]{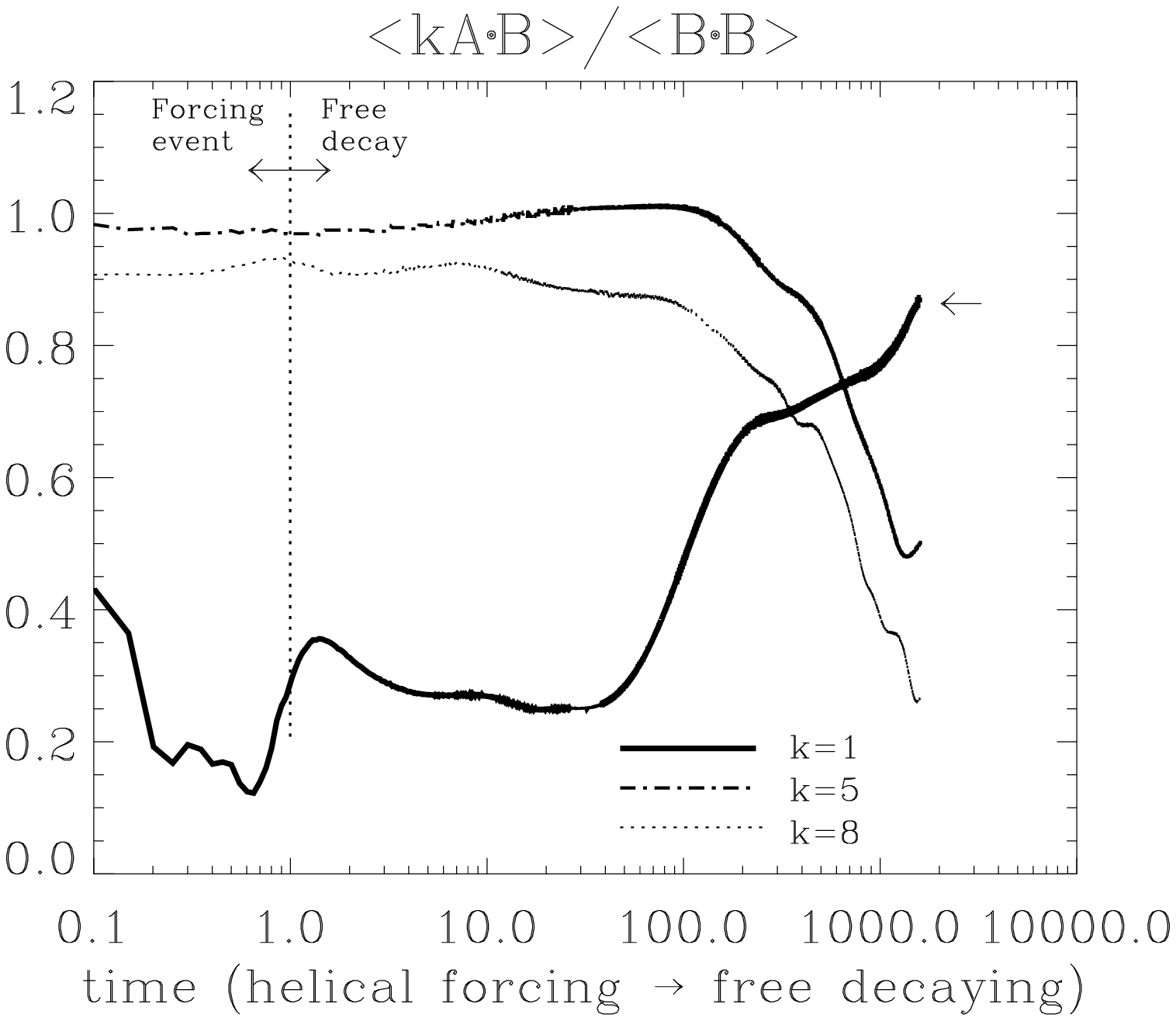}
     \label{f2b}
     }
  }
\caption{(a) Spectrum of decaying kinetic and magnetic energy. (b) Spectrum of decaying total magnetic energy $\langle B^2\rangle /2$ and helical magnetic energy $\langle k\mathbf{A}\cdot \mathbf{B} \rangle/2$. Statistically helicity is the correlation between different components of fields $\sim \Sigma_{i,\,j}^{i\neq j}\langle B_iB_j\rangle$. Magnetic energy includes helical and nonhelical component.
(c) Inverse cascade of total magnetic energy (solid red line), helical magnetic energy (dotted red line), and kinetic energy (dashed black line) of $k$=1, 5, 8. (d) Helicity ratio $kH_M/2E_M$.}
}
\end{figure}

\noindent Fig.\ref{f1a}, \ref{f1ab} show the decaying $E_V$ and (helical) $E_M$ in Fourier space. Fig.\ref{f2a} includes the evolving profiles of total $E_M$ ($B^2/2$, solid red line), helical $E_M$ ($\langle k\,\mathbf{A}\cdot \mathbf{B}\rangle/2$, dotted red line), and $E_V$ (dashed black line) in real space. Fig.\ref{f2b} shows the evolving helicity ratio `$kH_M/2E_M$' at $k=$1, 5, and 8. The conditions are the same as those of nonhelical case except that the system is initially driven by the helical magnetic energy. The helicity makes some distinct features discriminated from the nonhelical decaying turbulence. The migration of `$E_M$ peak' appears clearly, the strength of kinetic energy and overall decay rate of energy are lower than those of the nonhelical energy system. Like the nonhelical MHD system the magnetic energy is transferred to the kinetic eddy through Lorentz force. In principle the fully helical magnetic energy cannot be transferred to the kinetic eddies. However the unstable turbulent motion lacking in memory effect generates nonhelical component so that kinetic eddies receive energy from magnetic eddies. So although the strength of $E_V$ is weaker, there is limited contribution of $E_V$ to the induction of nonhelical magnetic field. Careful comparison of Fig.\ref{f2}, \ref{f2a}, \ref{f2b} shows that the kinetic energy is transferred toward both directions; consequently, $E_V$ in large and small scale leads to the forward and backward migration of $E_M$. In the early time regime ($1<t<\sim50$), the helicity ratio of magnetic energy keeps relatively constant in spite of the elevation of $E_M$. When the helicity ratio in large scale $E_M$ begins to accelerate ($t>\sim50$, Fig.\ref{f2b}), that of small scale $E_M$ starts falling. We can infer the transfer of total $E_M$ due to $E_V$ precedes that of helical $E_M$ due to the small scale helical magnetic field (`$\alpha$ effect). And then `$\alpha$ coefficient' interacts with large scale magnetic field directly leading to the cascade of small scale helical magnetic energy inversely. The role of pressure and advection term seems to be independent of the helicity ratio in the system.

\section{Theory: Models of energy transfer}
\subsection{Inverse cascade of helical magnetic energy}
As Fig.\ref{f1}-\ref{f2b} show, the magnetic energy in a decaying MHD system can be transferred toward the larger scale whether the field is helical or not. In case of helical field, toroidal magnetic field interacts with the helical fulid motion to generate (amplify) poloidal magnetic field, and the poloidal component generates (amplifies) the helical component through the interaction with plasma. The effect of helicity in small scale is represented by `$\alpha$', which can be considered as an independent coefficient in a homogeneous and isotropic (without reflection symmetry) system. According to the relative contribution to the field amplification, the helical dynamo is divided into `$\alpha^2$ dynamo', `$\alpha\Omega$ dynamo', `$\alpha^2\Omega$ dynamo' (Here `$\Omega$' indicates the differential rotation effect).\\

\noindent The large scale magnetic energy and helicity in a decaying system are represented by the solutions of coupled Eq.(\ref{Em with alpha coefficient}), (\ref{Hm with alpha coefficient}) \citep{Park 2014}:
\begin{eqnarray}
2\overline{H}_M(t)&=&(\overline{H}_M(0)+2\overline{E}_M(0))e^{2\int^t_0(\alpha-\beta-\eta)d\tau}
+(\overline{H}_M(0)-2\overline{E}_M(0))e^{-2\int^t_0(\alpha+\beta+\eta)d\tau},\label{HmSolution1}\\
4\overline{E}_M(t)&=&(\overline{H}_M(0)+2\overline{E}_M(0))e^{2\int^t_0(\alpha-\beta-\eta)d\tau}
-(\overline{H}_M(0)-2\overline{E}_M(0))e^{-2\int^t_0(\alpha+\beta+\eta)d\tau}.\label{EmSolution2}
\end{eqnarray}
(Here $\overline{E}_M(0)$ and $\overline{H}_M(0)$ are the initial large scale magnetic energy and helicity.)\\
\noindent With the positive initial magnetic helicity ($\alpha>0$), the first terms on the right hand side (RHS) are dominant. However, since there is no driving source, `$\alpha$ $(\sim \langle {\bf j}\cdot {\bf b}\rangle-\langle {\bf u}\cdot {\bf\omega}\rangle)$' and `$\beta$ $(\sim \langle u^2\rangle)$' eventually decay and converge to `$zero$'. But the combined index `$\alpha-\beta-\eta$' decreases to become negative with the finite diffusivity `$\eta$'. Consequently the strengths of (large scale) magnetic energy and helicity increase first, reach the peak, decay and converge to $zero$ eventually, which is consistent with the field evolution shown in Fig.\ref{f2a}. If the initial magnetic helicity in small scale is negative, i.e., $\alpha<0$ the second terms in RHS decide the profiles of $\overline{E}_M(t)$ and $\overline{H}_M(t)$ with the negative sign.\\

\subsection{Inverse transfer of nonhelical magnetic energy}
\noindent The inverse transfer of decaying nonhelical magnetic energy cannot be explained by Eq.(\ref{HmSolution1}), (\ref{EmSolution2}) because of the negligible `$\alpha$'. Instead there have been trials to explain the phenomenon qualitatively using the scaling invariant Navier Stokes and magnetic induction equation \citep{Olesen 1997, Ditlevsen et al. 2004} with numerical simulation \citep{Brandenburg et al 2014}. The incompressible free decaying MHD equations
\begin{eqnarray}
\frac{\partial {\bf u}}{\partial t}&=&-{\bf u}\cdot \nabla {\bf u}+\underbrace{{\bf b}\cdot \nabla {\bf b}-{\bf \nabla} P}_{\mathbf{j}\times \mathbf{b}-{\bf \nabla} p}+\nu{\bf \nabla}^2 {\bf u}\label{decaying momentum equation},\\
\frac{\partial {\bf b}}{\partial t}&=&\underbrace{\mathbf{b}\cdot \nabla \mathbf{u}-\mathbf{u}\cdot \nabla \mathbf{b}}_{\nabla \times \langle {\bf u} \times {\bf b} \rangle}+\eta\,{\bf \nabla}^2{\bf b}.
\label{decaying magnetic indution equation}
\end{eqnarray}
are invariant under the scaling transformation: $\mathbf{r}\rightarrow l\mathbf{r}$, $t\rightarrow l^{1-h}t$, $\mathbf{u}\rightarrow l^h\mathbf{u}$, $\nu\rightarrow l^{1+h}\nu$, $\mathbf{b}\rightarrow l^h\mathbf{b}$, $\eta\rightarrow l^{1+h}\eta$, $P\rightarrow l^{2h}P$, where `$l$' and `$h$' are arbitrary parameters. Then the scaled kinetic energy is
\begin{eqnarray}
\mathbb{E}_V(k/l, l^{1-h}t, Ll, K/l)&=&l^4 \frac{2\pi k^2}{(2\pi)^3}\int^L_{2\pi/K}d^3xd^3y\,\,e^{i\mathbf{k}\cdot (\mathbf{x}-\mathbf{y})}\langle \mathbf{u}(l\mathbf{x}, l^{1-h}t)\mathbf{u}(l\mathbf{y}, l^{1-h}t)\rangle\nonumber\\
&=& l^{4+2h}\mathbb{E}_V(k,t,L,K).\label{decaying E_V}
\label{decaying E_M}
\end{eqnarray}
Similarly the scaled magnetic energy is represented like `$\mathbb{E}_M(k/l, l^{1-h}t, Ll, K/l)=l^{4+2h}\mathbb{E}_M(k,t,L,K)$'. So the energy density spectrum is `$E_{V,\,M}(k/l, l^{1-h}t, lL, K/l)=l^{1+2h}E_{V,\,M}(k,t,L,K)$', which can be defined as `$E_{V,\,M}(k,\,t)=k^{-1-2h}\psi_{V,\,M} (k,\,t)$' with an arbitrary function `$\psi (k,\,t)$'
(\cite{Olesen 1997}, \cite{Ditlevsen et al. 2004}, references therein). These scaled energy density relations lead to
\begin{eqnarray}
\psi_{V,\,M}(k/l,\,l^{1-h}t)=\psi_{V,\,M} (k,\,t).
\label{psi relation}
\end{eqnarray}
If we differentiate Eq.(\ref{psi relation}) with respect to `$l$' and then set `$l=1$', we derive a differential equation:
\begin{eqnarray}
-k\frac{\partial \psi_{V,\,M}}{\partial k}+(1-h)t\frac{\partial \psi_{V,\,M}}{\partial t}=0.
\label{psi differential equation}
\end{eqnarray}
The general solution of this equation implies `$\psi_{V,\,M} (k,\,t)$' is the function of `$k^{1-h}t$', which implies the inverse transfer of energy with increasing time `$k\sim t^{1/(h-1)}$'. For example if the primordial energy is `$E_{V,\,M} (k,\,0)=k^{-1-2h}\,(q\equiv-1-2h)$', the energy at `$t$' is `$E_{V,\,M} (k,\,t)=k^{-1-2h}\psi_{V,\,M}(k^{1-h}t)=k^{q}\psi_{V,\,M}(k^{(3+q)/2}t)$'. More clearly, when $h<1$ (i.e., $q>-3$), the decaying energy can be transferred inversely. Briefly the energy migration is the result of scaled energy density relation, which is the intrinsic property of MHD equations. However this criterion is not valid in the whole range. The integration of $E_{V,\,M}\sim k^q\,\psi_{V,\,M}(k^{(3+q)/2}t)$ to get the total energy is known to yield an inconsistent result \citep{Ditlevsen et al. 2004}. The saturated energy spectrum of $E_V$ and $E_M$ in a `large $Pr_M$ MHD system' driven by the nonhelical kinetic energy has the relation of `$E_M^2=k^2E_V$' in the subviscous scale \citep{Park 2015}. When the background magnetic field `$b_{ext}$' is strong, turbulence due to the nonlinear effect becomes relatively weak compared with that of the guiding background magnetic field. This leads to the energy spectrum $E_V\sim k^{-4}$ and $E_M\sim k^{-1}$ \citep{Lazarian et al 2004}. If the forcing stops, $E_V$ should be cascaded toward small scale but $E_M$ should be transferred toward large scale. However $E_V$ and $E_M$ cannot migrate oppositely. With the strong $b_{ext}$ the assumption of self-similarity in the MHD equations is not valid. Moreover it is not yet clear if the scaled energy density relation can be applied to the helical field MHD system. More study is necessary.



\subsection{New model of energy transfer based on MHD equation}
We suggest a new dynamo model for the transfer of $E_M$. This model is not limited to the free decaying turbulence; rather, it covers the general mechanism of converting kinetic into magnetic energy and the local magnetic energy transfer. As mentioned dynamo is in fact the result of interaction between the velocity field `$\mathbf{u}$' and magnetic field `$\mathbf{b}$', i.e., EMF $\langle \mathbf{u} \times \mathbf{b}\rangle$ and cross helicity $\langle \mathbf{u} \cdot \mathbf{b}\rangle$. While EMF plays a role of explicit source of the magnetic field, the cross helicity constrains or suppresses the field profile implicitly \citep{Yokoi 2013}. In this paper we do not consider the effect of cross helicity; instead, we show how the curl of EMF $\nabla \times \langle \mathbf{u} \times \mathbf{b}\rangle$ and its decomposed terms `$\mathbf{b}\cdot \nabla \mathbf{u}$', `$-\mathbf{u}\cdot \nabla \mathbf{b}$' induce the magnetic field. The MHD system is filled with various kinds (scale, magnitude) of magnetic and velocity fields. If both velocity and magnetic field are composed of a single field vector, `$\nabla$' or `$\nabla\times$' is meaningless and no dynamo occurs. The fields should be at least locally inhomogeneous and anisotropic although the macroscopic system may be homogeneous and isotropic. So we consider two simplest cases that show the physical process of dynamo clearly: `$\mathbf{u}$' (or $\mathbf{b}$) is a single field vector, but `$\mathbf{b}$' (or $\mathbf{u}$) is composed of plural field vectors.\\

\begin{figure}
\centering{
  {
   \subfigure[]{
     \includegraphics[width=12cm]{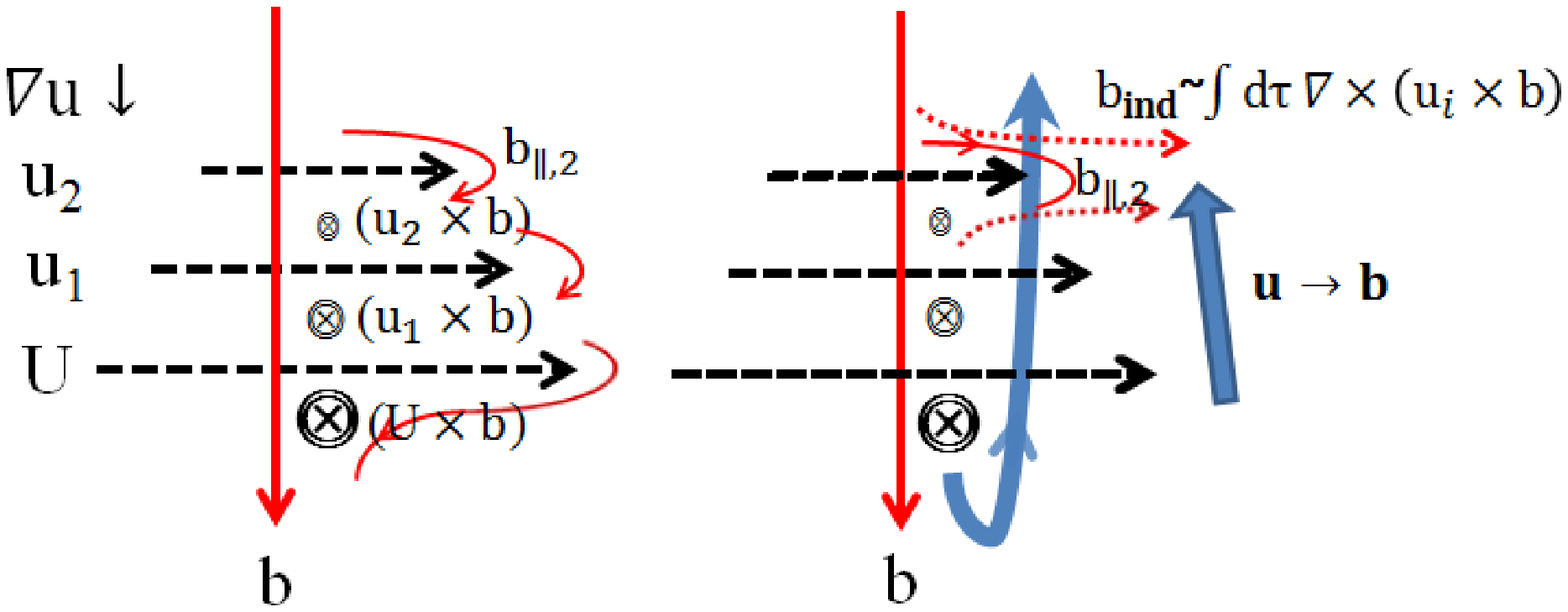}
     \label{f3}
   }\hspace{-0mm}
   \subfigure[]{
     \includegraphics[width=12cm]{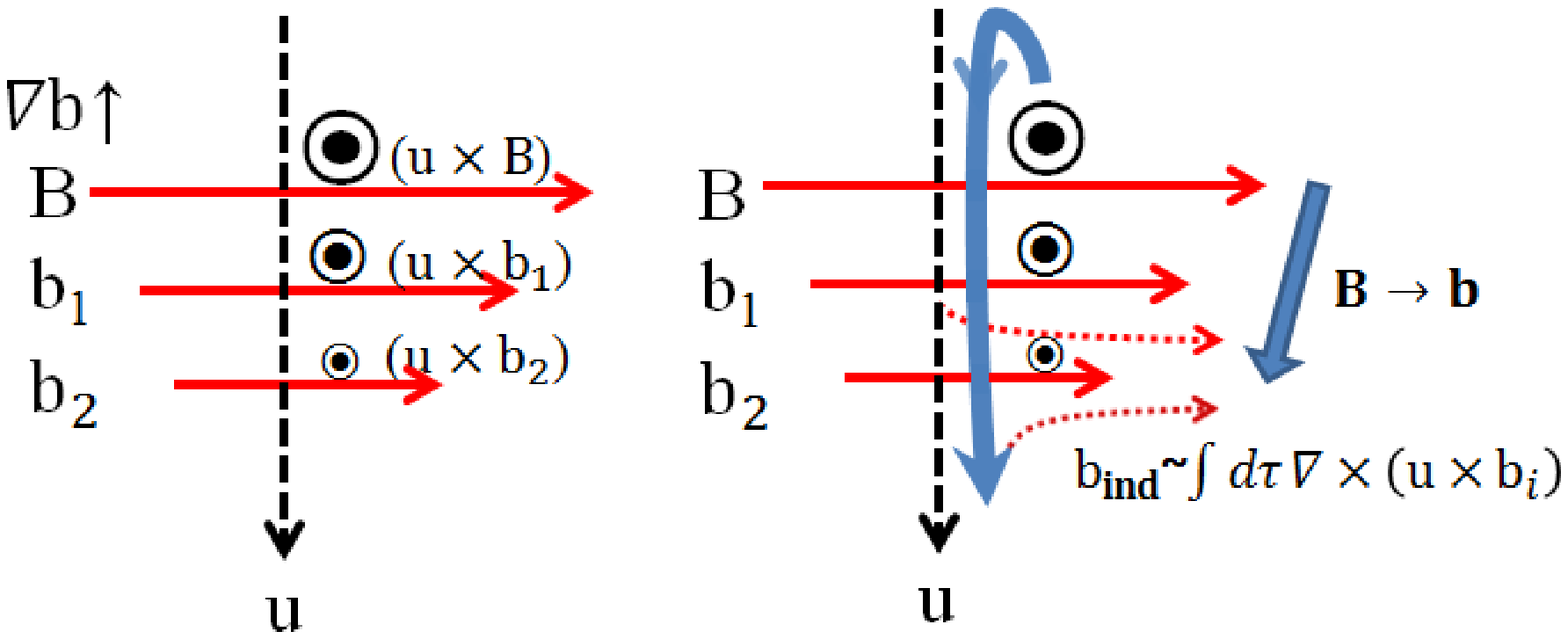}
     \label{f4}
     }
  }
\caption{(a) The energy in kinetic eddies is cascaded toward magnetic eddies in a nonlocal way. (b) The energy in a magnetic eddy is transferred toward its adjacent eddy.}
}
\end{figure}

\noindent In Fig.\ref{f3} we assume the magnetic field heads for the gradient of velocity field. That is, we set the magnetic field: `$\mathbf{b}=(0,\,0,\,b)$' and velocity field: `$\mathbf{u}_i=(0, \,u_i(z), \,0)$'. The strengths of velocity fields `$\mathbf{u}_i$ ($i=1,\,2$)' and `$\mathbf{U}$' are assumed to be the function of `$z$' in order of $u_2<u_1<U$. Also the velocity field is not too strong compared with `$b$'. Then we can make clear the meaning of first term `$\mathbf{b}\cdot \nabla \mathbf{u}$' in Eq.(\ref{decaying magnetic indution equation}). Mathematically it is the contraction of second order tensor into the first order one, i.e., vector:
\begin{eqnarray}
\mathbf{b}\cdot \nabla \mathbf{u}\rightarrow b\,\hat{z}\cdot \bigg(\hat{x}\frac{\partial}{\partial x}+\hat{y}\frac{\partial}{\partial y}+ \hat{z}\frac{\partial}{\partial z}\bigg)u_i(z)\,\hat{y}=b\,\frac{\partial u_i(z)}{\partial z}\,\hat{y}.
\label{b_cdot_nabla_u}
\end{eqnarray}
Since strength of `$\mathbf{u}_i(z)$' increases toward `$\hat{z}$', the induced magnetic field `$\mathbf{b}_{ind}$' is parallel to `$\mathbf{u}_i$'. And `$\mathbf{b}_{ind}$' is merged with `$\mathbf{b}_{\|}$' which is partially dragged `$\mathbf{b}$' by the fluid motion `$\mathbf{u}_i$'. If we assume `$\mathbf{b}_\|$' is the strongest at `$U$' and `$\partial u_i/\partial z$' is uniform, the transferred magnetic field `$b_{ind}+b_\|$' will have the largest value at `$U$'. However since `$\mathbf{b}\cdot \nabla \mathbf{u}$' is the part of $\nabla \times (\mathbf{u}\times \mathbf{b})$, some physical information is missing. So we need to derive the induction of magnetic field through the curl of EMF directly. As mentioned `$\mathbf{b}$' interacts with the fluid `$\mathbf{u}_i$' to generate EMF varying from the smallest $\langle \mathbf{u}_2\times \mathbf{b} \rangle$ to the largest $\langle \mathbf{U}\times \mathbf{b} \rangle$. These differential magnitudes in EMF yield the clockwise rotational effect $\nabla \times \langle \mathbf{u}_i\times \mathbf{b}\rangle>0$, which induces the magnetic field $\mathbf{b}_{ind}\sim \int d\tau \nabla \times \langle \mathbf{u}_i\times \mathbf{b}\rangle$. Therefore, $\mathbf{b}_{ind}$ and $\mathbf{b}_{\|,\,\,i}$ are in the complementary relation to cause the growth of net magnetic field. If the strength of `$\mathbf{b}_\|$' is assumed to be uniform, $b_{ind}+b_{\|,\,\,i}$ will be the strongest at $u_2$. However if `$b_\|$' is proportional to the strength of `$\mathbf{u}_i$', the combined magnetic field will have the largest value between `$\mathbf{u}_i$' and `$\mathbf{U}$'. This can be one of the reasons why the peak of $E_M$ in a non helically driven MHD system (small scale dynamo) is located between the injection and viscous scale. Of course the influence of eddy turnover time `$\tau_i$' and dissipation effect `$\sim k^2\,b_i$' need to be considered in order to explain the reason more exactly. On the other hand if the direction of $\mathbf{b}$ is reversed ($\mathbf{b}\cdot \nabla \mathbf{u}<0$), $\mathbf{b}_{ind}$ is in the opposite direction of $\mathbf{b}_\|$ resulting in the dissipation of net magnetic field. This is the process of nonlocal energy transfer from $E_V$ to $E_M$.\\

\noindent In the same way, we can explain the local energy transfer in magnetic eddies as shown in Fig.\ref{f4}. At this time the fluid `$\mathbf{u}$' is heading for the decreasing $equi$-magnetic field line ($\mathbf{u}\cdot \nabla \mathbf{b}<0$). Each velocity and magnetic field can be represented like `$\mathbf{u}=(0,\,0,\,u)$' and `$\mathbf{b}_i=(0, \,b_i(z), \,0)$'. The strength of magnetic field is the function of `$z$' in the order of $b_2<b_1<B$ and $\partial b_i(z)/\partial z < 0$. Then the second term `$-\mathbf{u}\cdot \nabla \mathbf{b}_i$' in Eq.(\ref{decaying magnetic indution equation}) is
\begin{eqnarray}
-u\,\hat{z}\cdot \bigg(\hat{x}\frac{\partial}{\partial x}+\hat{y}\frac{\partial}{\partial y}+ \hat{z}\frac{\partial}{\partial z}\bigg)b_i(z)\,\hat{y}=-u\,\frac{\partial b_i(z)}{\partial z}\,\hat{y}\,\,\rightarrow \,\,\bigg|u\,\frac{\partial b_i(z)}{\partial z}\bigg|\,\hat{y}.
\label{b_cdot_nabla_u}
\end{eqnarray}
So the direction of induced magnetic field `$\mathbf{b}_{ind}$' is $\hat{y}$, which is parallel to `$\mathbf{b}_i$'. Also their summation is the largest at $\mathbf{B}$. However, this cannot explain local energy transfer. So we need to derive the induced magnetic field using the mathematical definition of curl operator again. The interaction between `$\mathbf{u}$' and `$\mathbf{b}$' generates EMF varying from the smallest $\langle \mathbf{u}\times \mathbf{b}_2\rangle$ to the largest $\langle \mathbf{u}\times \mathbf{B} \rangle$. Their differential strengths create the rotational effect, which induces magnetic field $\mathbf{b}_{ind}\sim \int d\tau \nabla \times \langle \mathbf{u}\times \mathbf{b}_i\rangle$. This induced `$\mathbf{b}_{ind}$' is parallel to `$\mathbf{b}_i$' and largest at `$\mathbf{b}_2$' indicating the energy migration from the strongest magnetic field to the weakest one, which is independent of an eddy scale. This explains the inverse transfer of $E_M$ in a decaying MHD system. Like the nonlocal energy transfer the direction of local $E_M$ transfer is determined by the location of velocity field, relative strength of energy among eddies, eddy turnover time, dissipation effect. Also we can see that `-$\mathbf{b}_{ind}$' generated by the oppositely directed `$\mathbf{u}$' annihilates `$\mathbf{b}_i$'.\\

\noindent The induced magnetic field also constrains the fluid motion. The current density `$\mathbf{J}$' in Lorentz force (Eq.\ref{decaying momentum equation}) is derived from the flow of electric charge `$q\mathbf{u}$' carried by the fluids. In view of macroscopic fluid model, `$J$' is replaced by $\nabla \times B$, meaning the induced magnetic field is also able to constrain the fluid motion. In Fig.\ref{f3}, where the magnetic field `$b\,\hat{z}$' crosses the velocity fields, the directions of $\mathbf{b}_{ind}$ and $\nabla \times \mathbf{b}_{ind}$ are `$\hat{y}$' and `$\hat{x}$' respectively, the direction of $(\nabla \times \mathbf{b}_{ind})\times \mathbf{b}_{ind}$ is `$\hat{z}$'. This appears to perturb the fluids `$\mathbf{u}_i$' perpendicular. However, $(\nabla \times \mathbf{b}_{ind})\times \mathbf{b}_{ind}$ can be decomposed into the magnetic pressure `$-\nabla b^2_{ind}/2$' and magnetic tension `$\mathbf{b}_{ind}\cdot \nabla \mathbf{b}_{ind}$'. Magnetic tension cancels the magnetic pressure parallel to the field so that `$\mathbf{j}_{ind}\times \mathbf{b}_{ind}$' seems to press the fluid perpendicular.
However, this tension is actually $\mathbf{\hat{b}}_{ind}\mathbf{\hat{b}}_{ind}\cdot \nabla b^2_{ind}/2 + b^2_{ind}\mathbf{\hat{b}}_{ind}\cdot \nabla \mathbf{\hat{b}}_{ind}=\mathbf{\hat{b}}_{ind}\nabla_\| b^2_{ind}/2 + b^2_{ind}\mathbf{\hat{\kappa}}$ where `$\mathbf{\hat{\kappa}}$' is a measure of the curvature of $\mathbf{b}_{ind}$. So the exact Lorentz force is $-\nabla_\bot b^2_{ind}/2 + b^2_{ind}\mathbf{\hat{\kappa}}$. This curvature related force gives energy to the fluid motion. Thermal and magnetic pressure suppress the fluid motion, but magnetic tension boosts it. On the other hand in Fig.\ref{f4}, $(\nabla \times \mathbf{b}_{ind})\times\mathbf{b}_{ind}$ perturbs the fluid motion.\\

\noindent These two cases explain the amplification, more exactly migration of $E_M$ and its constraint on $E_V$. Real distribution of $E_V$ and $E_M$ is more complicated, but basically they can be replaced by the combination of these two structures. What we have neglected is the effect of cross helicity $\langle \mathbf{u}\cdot \mathbf{b}\rangle$. When magnetic field `$\mathbf{b}$' is (anti) parallel to the fluid motion `$\mathbf{u}$', EMF is $zero$ so that `$\mathbf{b}_{ind}$' is not generated. But more detailed stochastic analysis shows the effect of cross helicity is generated when the fourth order moment is decomposed into the combination of second order one to close the MHD equations: $\langle uubb_{ind}\rangle\simeq \langle uu\rangle\langle bb_{ind}\rangle+\langle ub\rangle\langle ub_{ind}\rangle$ (Quasi Normal approximation, \cite{Kraichnan and Nagarajan 1967}, \cite{Park 2015}), which qualitatively matches the cross helicity term `$\langle \mathbf{u}_i\cdot \mathbf{b}_{ind}\rangle$' shown in Fig.\ref{f3} or \ref{f4}. For detailed investigation of cross helicity and anisotropy in the magnetized plasma system, more elaborate method is required. But we will not discuss about those topics here.

\section{Summary}
We have shown the inverse transfer of $E_M$ in a free decaying MHD turbulence. In the typical 2D hydrodynamic and 3D MHD turbulence system, the inverse transfer of energy is related to the conservation of physical invariants such as energy, enstrophy, or magnetic helicity. The overall mechanism of inverse cascade of helical $E_M$ is well understood. However it is still tentative if the magnetic helicity is prerequisite to LSD. Moreover since the helicity is negligible or completely absent from some celestial objects that are filled with various scales of magnetic fields, the exact comprehension of (inverse) transfer of general $E_M$ is important to understand the origin and mechanism of magnetic field evolution in the present universe.\\

\noindent As we have seen, the inverse transfer of nonhelical $E_M$ is the result of intrinsic property of scaling invariant MHD equations. The evolving energy spectrum at `$t$' is represented by the initial energy distribution $E_{V,\,M}(k,\,0)\sim k^q$ and $\psi(k^{(3+q)/2}t)$. However this relation, limited to the local range, may induce some inconsistent inference for the direction of energy transfer.\\

\noindent We introduced a dynamo model based on the magnetic induction equation $\partial \mathbf{b}/\partial t \sim \nabla \times (\mathbf{u}\times \mathbf{b})\sim \mathbf{b}\cdot \nabla \mathbf{u}-\mathbf{u}\cdot \nabla \mathbf{b}$. This model is not limited to the inverse transfer of decaying $E_M$, but it explains the general mechanism of $E_M$ transfer. The migration of $E_M$ is in fact the continuous induction of magnetic field guided by $E_V$. The energy in the kinetic eddies is non locally transferred to the magnetic ones through `$\mathbf{b}\cdot \nabla \mathbf{u}$'; and, the energy in the magnetic eddies is locally transferred to their adjacent ones through `$-\mathbf{u}\cdot \nabla \mathbf{b}$'. As the model shows, what decides the direction of magnetic energy transfer is not the eddy scale size, but the guide of velocity field and relative energy difference between eddies. In a mechanically driven system, the pressure and advection term transfer $E_V$ chiefly toward the smaller scale whose eddy turnover time decreases by $\sim 1/k$ but dissipation effect elevates by $\sim k^2$. Then $E_V$ cascaded to the small scale induces $E_M$ continuously forming the peak of $E_M$ in small scale. But strictly speaking energy migrates toward both directions. If the forcing in the system stops, $E_V$ in small scale fades away more quickly than that of large scale. Consequently $E_V$ in large scale generates $E_M$ in large scale and decides the profile of energy spectrum in the system. Again magnetic back reaction due to Lorentz force generated by this $E_M$ constrains (suppresses or boosts) the fluid motion, which is consistent with other numerical results.\\

\noindent In addition to this current model and numerical simulation, a more detailed analytic theory and various numerical simulations with arbitrary helicity ratio are required. Since we need to trace the evolution of $E_V$ and $E_M$ together, Eddy Damped Quasi Normalized Markovianized (EDQNM) approximation is a more suitable method than other stochastic models \citep{Kraichnan and Nagarajan 1967, Pouquet et al 1976, Park 2015}. According to the numerical test of EDQNM for a unit $Pr_M$ MHD system\citep{Son  1999}, the spectrum of low $k$ is not modified, but the peaks of $E_V$ and $E_M$ clearly migrates toward large scale. We think the approximation of EDQNM with large $Pr_M$ can show clearer transfer of $E_M$ toward the large scale and give us the hints about the exact representation of $\psi_{V,\,M}(k,\,t)$ beyond the inertial range. We leave these topics for the future work.


\section{acknowledgement}
KP acknowledges support from the National Research Foundation of Korea through grant
2007-0093860.

\end{document}